\documentclass[10pt,letterpaper,twocolumn]{article} 

\usepackage[usenames,dvipsnames]{color}
\usepackage{ol2}
\usepackage{amsmath,array,amssymb}
\usepackage{graphicx}
\usepackage{hyperref}
\usepackage{textcomp}
\usepackage{braket}
\usepackage{multirow}
\usepackage{indentfirst}
\usepackage[FIGTOPCAP,raggedright,nooneline,bf,footnotesize]{subfigure}
\usepackage{paralist}
\usepackage{overpic}

\newcommand{\figref}[1]{Fig.~\ref{#1}}

\newcommand{\corf}{r}

\newcommand{\sep}{\sigma_{\text{}}}

\begin{document}
\twocolumn[ 
\title{Demonstration of spectral correlation control in a source of polarization entangled photon pairs at telecom wavelength}
\author{Thomas Lutz,$^{1,2,*}$ Piotr Kolenderski,$^{1,3,*}$ and Thomas Jennewein$^{1}$}
\address{
$^1$Institute for Quantum Computing, University of Waterloo, 200 University Ave.~West, Waterloo, Ontario, Canada, N2L 3G1
\\
$^2$Institut f\"ur Quantenmaterie, Universit\"at Ulm, 89069 Ulm, Germany \\
$^3$Faculty of Physics, Astronomy and Informatics, Nicolaus Copernicus University, Grudziadzka 5, 87-100 Torun, Poland\\
$^*$Corresponding author: tlutz@uwaterloo.ca, kolenderski@fizyka.umk.pl
}

\begin{abstract}
Spectrally correlated photon pairs can be used to improve performance of long range fiber based quantum communication protocols. We present a source based on spontaneous parametric down-conversion producing polarization entangled photons without spectral filtering. In addition, the spectral correlation within the photon pair can be controlled by changing the pump pulse duration or coupled spatial modes characteristics. The spectral and polarization correlations were characterized. The generated photon pairs feature both positive spectral correlations, no correlations, or negative correlations and polarization entanglement with the fidelity as high as 0.97 (no background subtraction) with the expected Bell state.
\end{abstract}

] 

\maketitle

Controlling the spectral correlations of a polarization entangled photon pair produced via spontaneous parametric down-conversion (SPDC) could have important benefits to applications in optical
quantum information. Photonic quantum gates require pure states, which can be created by heralded sources producing pairs of spectrally decorrelated photons \cite{Uren2005,Uren2007,Osorio2008,Mosley2008,Kolenderski2009,Eckstein2011,Evans2010,Gerrits2011,Hendrych2007,Valencia2007,Jin2013}. On the other hand, long distance fiber based quantum communication and quantum metrology \cite{Dayan2005,Dayan2007} suffers from chromatic dispersion, which could potentially be improved \cite{Lutzthes2013} with positive spectral correlations\cite{Kuzucu2005,Kuzucu2008,Harder2013}.
Here we experimentally demonstrate the effective control of spectral correlations in a photon pair source, based on a $\beta-$barium borate (BBO) crystal which we characterized previously in Ref.~\cite{Lutz2013}. It produces polarization entangled pairs in the telecom band and can be tuned to create negative, none or positive spectral correlations. In addition to that we show high quality polarization entanglement without using spectral filtering or compensation crystals by adopting the compensation scheme shown in Ref.~\cite{Kim2002}.

\begin{figure}[h]
\centering
\includegraphics[width=0.8\columnwidth]{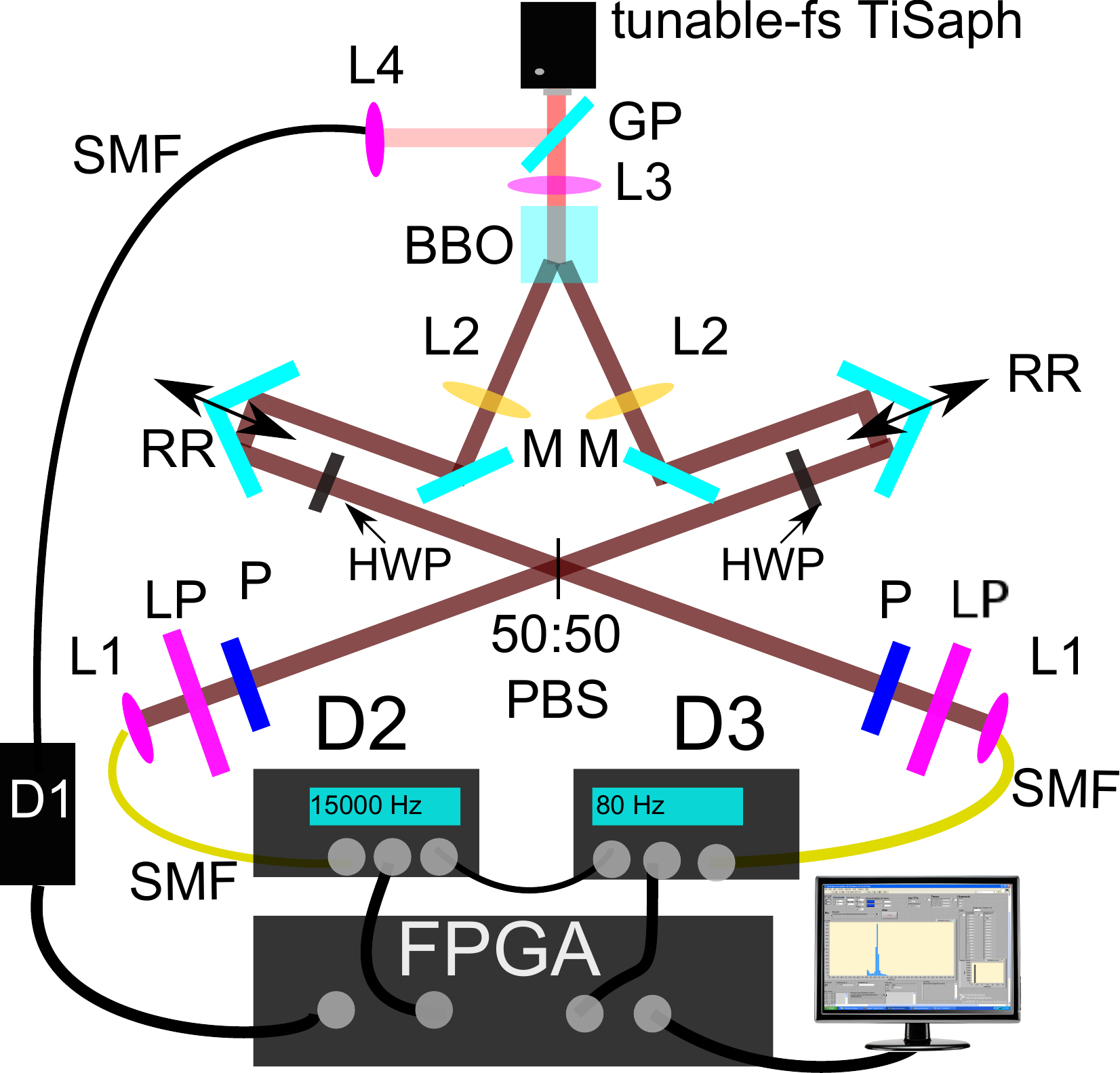}
\caption{The experimental setup. The BBO crystal is pumped using a Ti:Sapphire laser. The beam is focused onto the crystal by a lens (L3). Down-converted photons are collimated using lenses (L2, f=$25$ cm).  The retroreflectors are used to adjust the path lengths of the generated photons. A half wave plate (HWP) rotates the polarization by 90 degrees which results in splitting both photons at the polarizing beam splitter (PBS).  After they pass polarizers (P) they are coupled into single mode fiber (SMF-28e) by aspheric lenses (L1,f=15.4 mm). The stray light is filtered out by a longpass filter (Semrock BLP01-1319R-25). Photons are detected using two InGaAs/InP detectors D2 and D3. Timing analysis is performed using FPGA electronics. A small percentage of the pump light is directed to a photo diode (DET10A) D1 using a glass plate (GP), a coupling lens (L4) and a SMF at 800 nm. D1 is used to measure pump pulse time. }
\label{fig:exp-schem}
\end{figure}

In the SPDC process, one photon of the pump converts into a photon pair. Energy and momentum conservation relations, jointly described as phase matching,  and the properties of the pump photons, govern the characteristics of the generated photons. The probability amplitude for a photon pair emission in a given direction and at a given frequency can be described by the product of the pump spatio-temporal amplitude and the phase matching function \cite{Kim2002,Kolenderski2009}. The phase matching depending on the properties of the nonlinear media specifies the allowed emissions. Typically, the output photons are coupled into optical fibers. This corresponds to collecting photons from a specific range of directions, defined by the fiber and the optics. From this point of view, coupling can be understood as an additional condition to phase matching. Therefore, one can introduce an effective phase matching function (EPMF) \cite{Kolenderski2009,Lutz2013}, which fully describes the joint effect of the crystal and coupling into fibers. We approximate the EPMF using a Gaussian function with the characteristic width $\sep$. The spectral part of the pump can also be approximated by a Gaussian function with the characteristic spectral width $1/(2\tau_{\text{pump}})$. The spectral wave function then reads:
\begin{equation}
\psi(\nu_s,\nu_i)=N\exp\left(-\frac{\left(\nu _s-\nu _i\right){}^2}{\sep^2}\right)\exp\left(-\frac{\left(\nu _s+\nu _i\right)^2\tau_p^2}{4}\right).
\label{eq:wf}
\end{equation}

It has already been show experimentally \cite{Lutz2013} that the EPMF of the type II BBO crystal cut for telecom wavelength facilitates spectral correlation control. Tuning capabilities arise due to the variable pulse duration $\tau_p$ of the pump laser and the variable collection optics which influences the characteristic width $\sep$ of the EPMF. We introduce a spectral correlation parameter defined as $\corf=\frac{<\nu_s \nu_i>}{\sqrt{<\nu_s^2> <\nu_i^2>}}$, which for the approximate wave function introduced in \eqref{eq:wf}, can be computed as:
\begin{equation}
\corf=\frac{1/4\tau_p^2-\sep^2}{1/4\tau_p^2+\sep^2}.
\label{eq:corf}
\end{equation}

The source allows to generate polarization entangled pairs with the desired spectral correlations. In order to ensure the quality of  polarization entanglement, we use a technique previously described in Refs.~\cite{Kim2002,Poh2007}. The polarization of one  photon from the pair is rotated by 90 degree before combining them on a polarizing beamsplitter (PBS) and being coupled into single mode fibers (SMF). Using this approach, no compensating crystals are required. In addition, this allows to avoid problems arising due to the spectral distinguishably of the extraordinary and ordinary photons as those are coupled in to two distinct fibers. This way we do not have to use spectral filtering to improve polarization entanglement visibility.

The aim of the experiment is to characterize the spectral correlation and polarization entanglement of photon pairs. The schematic setup of the experiment is shown in \figref{fig:exp-schem}. The crystal used in this source is a BBO cut at $29.14$ degree for type II phase matching. A pump photon at wavelength $775$ nm is converted into a photon pair at wavelength $1550$ nm.  The crystal is pumped using either a CW (MIRA-900,Coherent) or a femtosecond (TiF-50M, Atseva) tunable Ti:Sapphire laser.  The generated photons are coupled into single mode fibers (SMF-28e, Corning) using collection optics. The pump spatial mode diameter was 61 $\mu$m and the collection modes are 420 $\mu$m. The down converted photons are then detected with a combination of two detectors: (D2) MPD InGaAs/InP single photon avalanche diode \cite{Tosi2012a}, which gates (D3) idQuantique id201 InGaAs/InP detector. A fast photodiode DET10A (D1) is used to measure the pump pulse reference time and to gate D2 detector. Photon pairs and pulse arrival times are registered using a programmable FPGA (UQDevices).\\ 

\begin{figure}[t]
\centering
\begin{tabular}{c c c}
\subfigure[$\tau_p=70$ fs]{\includegraphics[width=0.3\columnwidth]{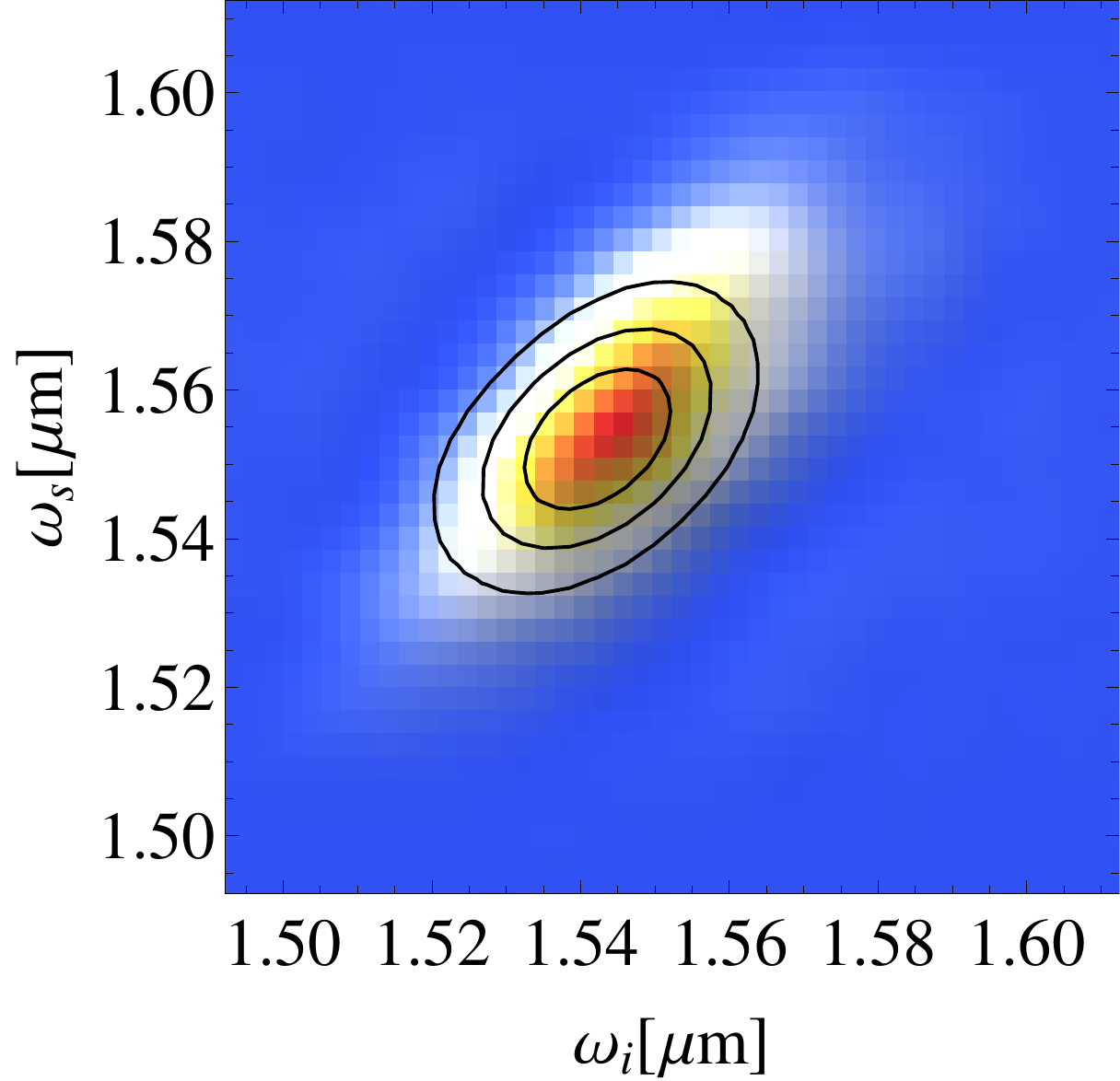}} &
\subfigure[$\tau_p=98$ fs]{\includegraphics[width=0.3\columnwidth]{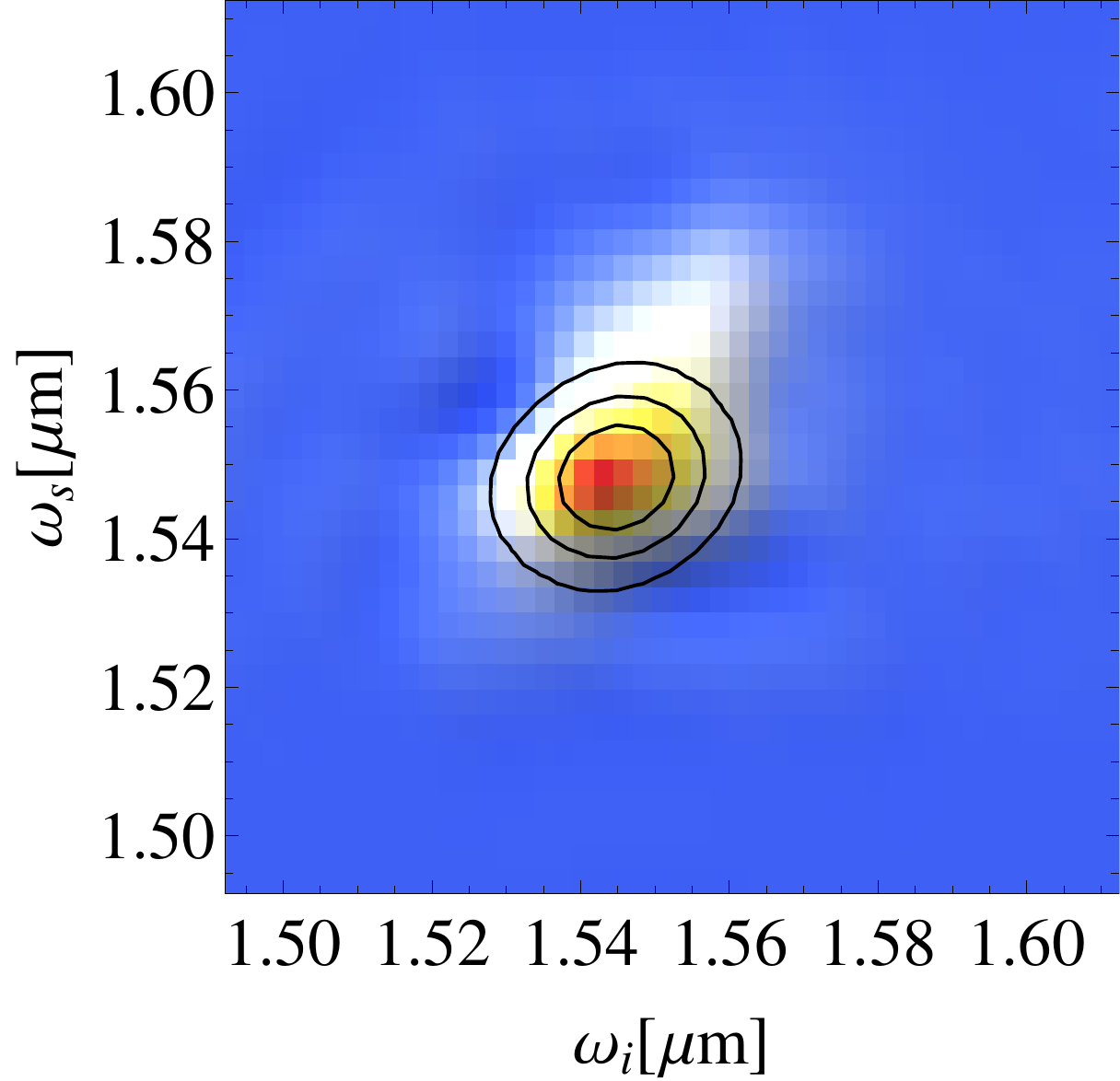}} &
\subfigure[CW]{\includegraphics[width=0.3\columnwidth]{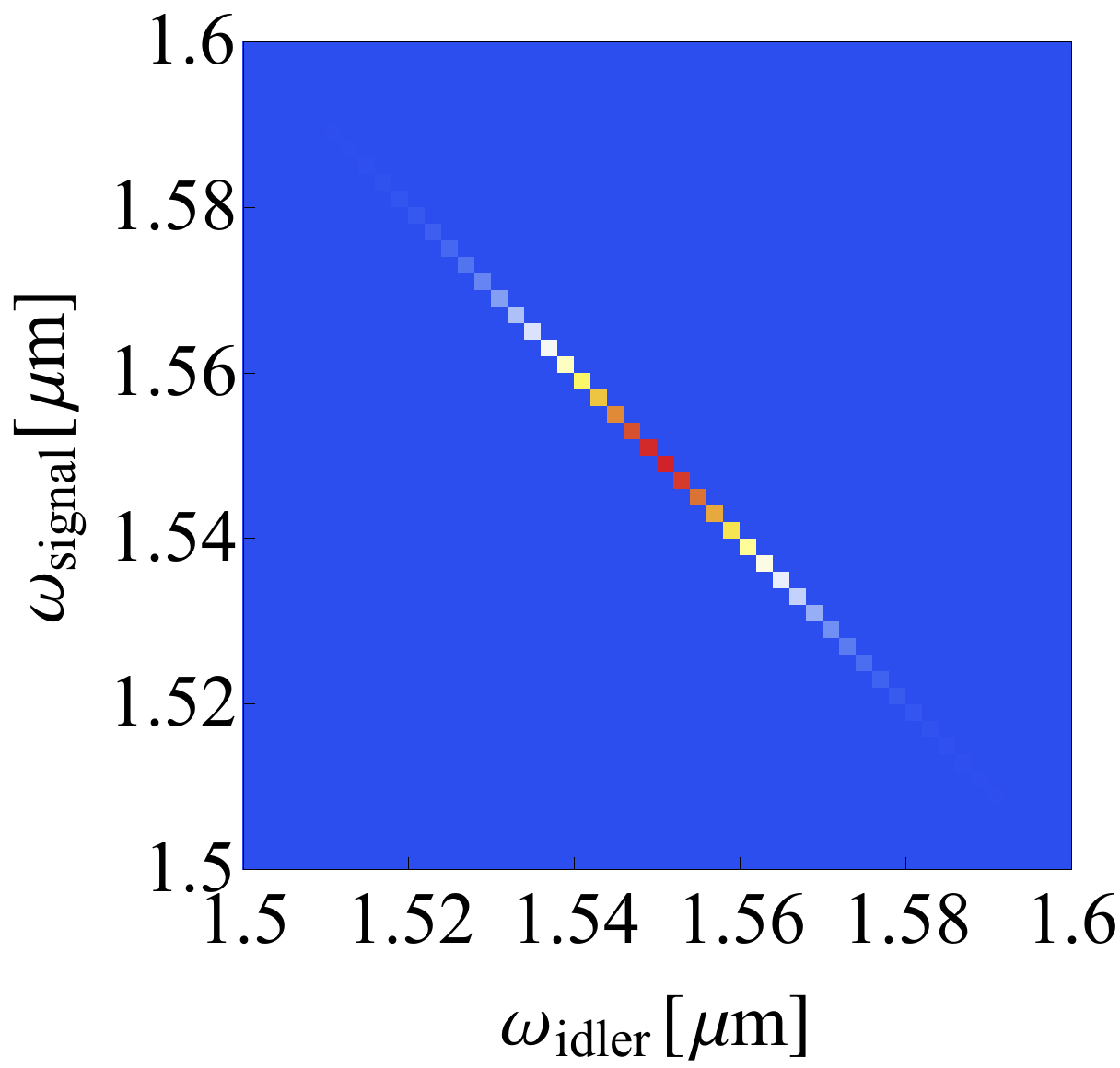}} \\
\end{tabular}
\caption{Measured joint spectral functions for pump durations of  (a) $70$ fs, (b) $98$ fs.  (c) Measurement taken with a continuous pump laser (CW) pump. The contour lines correspond to the theoretical prediction.}
\label{fig:jointspectra}
\end{figure}

First we investigate the spectral correlation within a photon pair. Based on our theoretical predictions \cite{Kolenderski2009}, the source is expected to generate photons featuring spectral correlations which are: a) positive when $\tau_p<110$ fs, b) reduced for $\tau_p \approx 110$ fs and c) negative when $\tau_p > 110$ fs. We perform our measurements for three pump settings: $\tau_p=70$ fs, $98$ fs and CW. We use a  method similar to the one already used to reconstruct EPMF in Ref.~\cite{Lutz2013}. The results are presented in \figref{fig:jointspectra} and Tab.~\ref{tab:bell}. It can be seen in \figref{fig:jointspectra}  that the joint-spectrum  is (a) spectrally positively correlated for $\tau_p=70$ fs, (b) shows very little correlations for $\tau_p=98$ fs, and (c) is spectrally anticorrelated in case of CW pumping.  In order to directly relate the experimental results with our theory, we compare the correlation factors defined by \eqref{eq:corf} and show the theoretical prediction for the joint spectrum as contours in \figref{fig:jointspectra}. The results for the correlation parameter, r, can be seen in Tab.~\ref{tab:bell}.


For the first two measurements we use a fast photodiode D1 to measure the pump pulse arrival time. This, in combination with the relative detection time of a generated photon pair allows us to reconstruct the spectral correlation characteristics \cite{Lutz2013}. The spectral resolution is limited by the timing jitter of the single photon detectors and photodiode. In order to improve the final results we use a deconvolution technique based on the overall timing resolution. Here we measured the following time jitters:  $43$ ps for the photodiode, $127$ ps for MPD detector and $400$ ps for id201 detector. The finite resolution  and experimental uncertainties are  reasons for the slight difference between predicted and measured correlation parameters for pulsed pump settings, see Tab.~\ref{tab:bell}. For the CW setting, our method of post-processing assumes perfect monochromatic pump laser, which in combination with energy conservation results in perfect anticorrelation.

\begin{table}
\centering
\begin{tabular}{c | c | c  |c |c |c}
$\tau_p$[fs]& r & S & vis  & fidelity & concur.\\
\hline\hline
70&  0.55  & $2.19 \pm 0.1.$ & 99 / 87  & 0.97 & 0.75 \\
98& -0.05  & $2.37 \pm 0.05$ & 99 / 86   & 0.94 & 0.76\\
CW& -1  & $2.42 \pm 0.20$ & 98 / 84 &  n/a & n/a \\
\end{tabular}
\caption{Measured spectral and polarization characteristics of photon pairs generated by the source. See text for a description of the parameters.}
\label{tab:bell}
\end{table}

\begin{figure}[h!]
\centering
\begin{tabular}{c c}
\subfigure[HV]{\includegraphics[width=0.45\columnwidth]{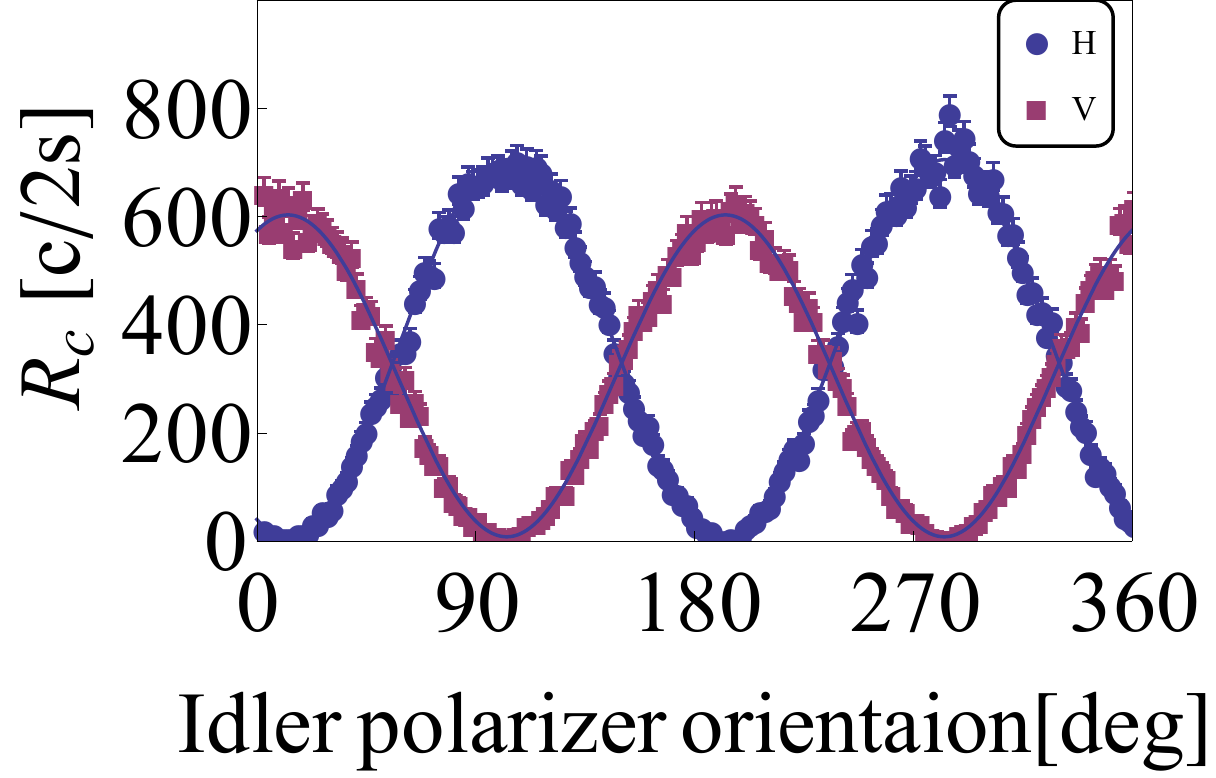}} &
\subfigure[DA]{\includegraphics[width=0.45\columnwidth]{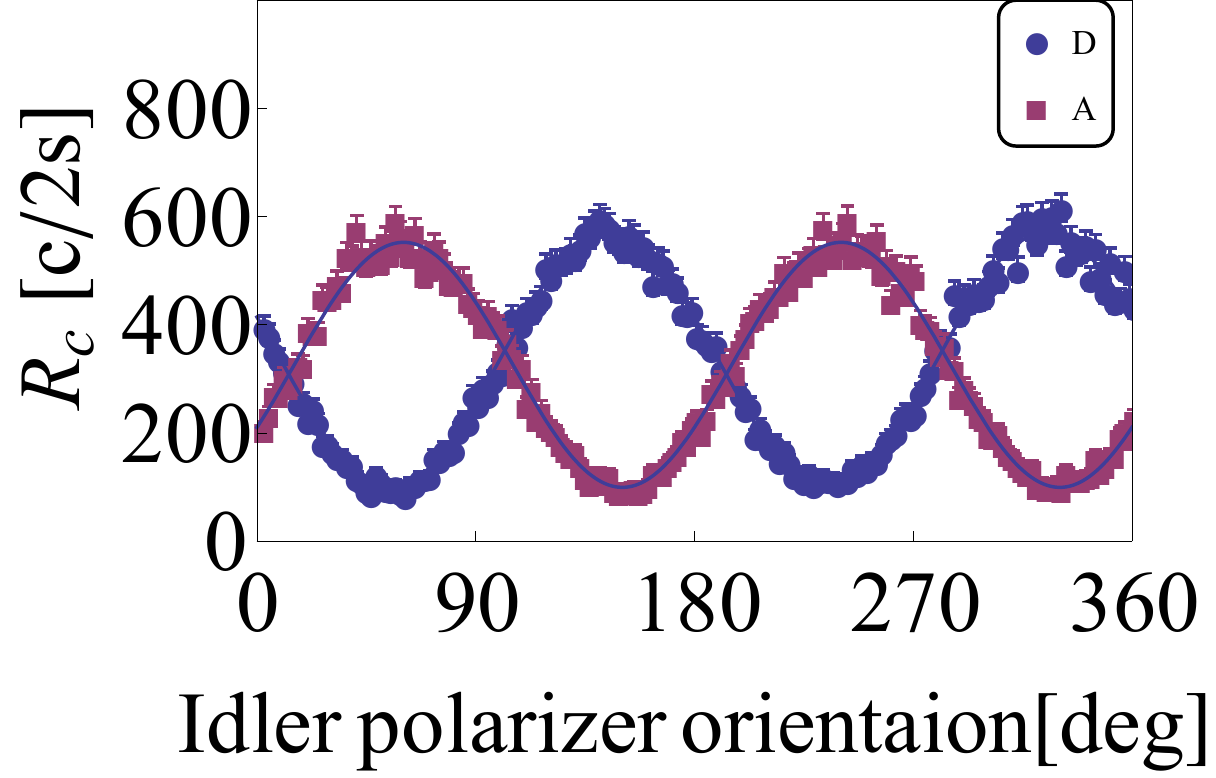}}\\
\subfigure[Re]{\includegraphics[width=0.45\columnwidth]{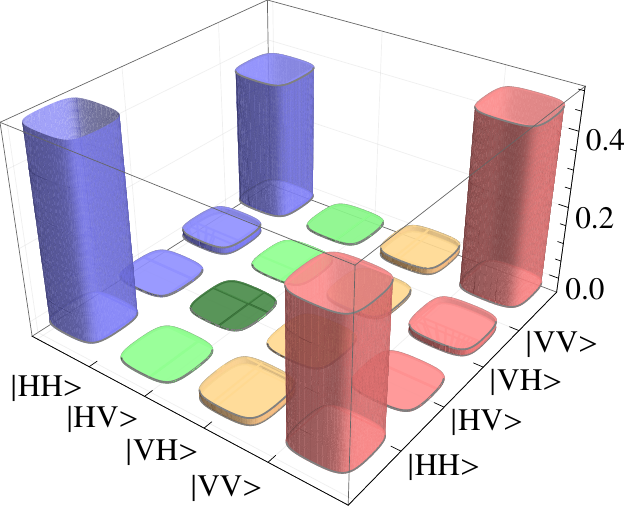}}&
\subfigure[Im]{\includegraphics[width=0.45\columnwidth]{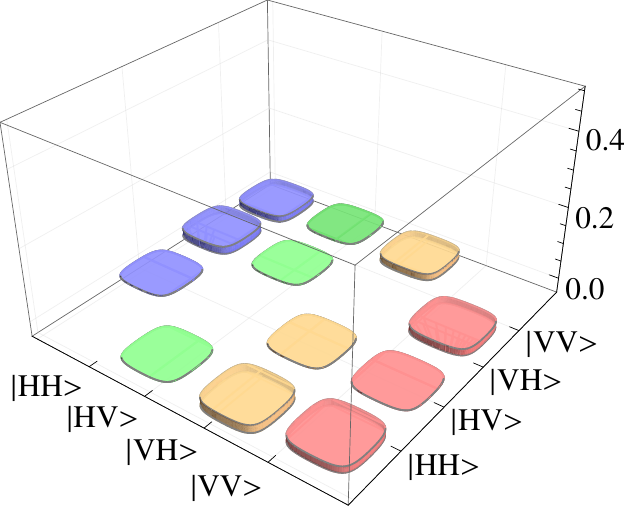}}

\end{tabular}
\caption{The measured polarization entanglement interference fringes in (a) HV and (b) AD basis for a pulse duration of $\tau_p=98$ fs. (c),(d) Tomographically reconstructed polarization state produced by the source when pumped with $\tau_p=98$ fs pulses. The fidelity between the measured state and the expected entangled state $(\ket{HH}+\ket{VV})/\sqrt{2}$ is $0.94$  (without background subtraction). The entanglement monotone (concurrence) is found to be $0.75$. The purity of the reconstructed state is $0.90$. }
\label{fig:fringes}
\end{figure}

Now, we move on to polarization entanglement characterization. We measure polarization interference fringes by fixing the signal polarizer orientation in the arm monitored by MPD detector to H, V, D or A polarization and count coincidences in the arm monitored by id201 detector for the full range of idler polarizer orientations. \figref{fig:fringes} shows the resulting interference fringes in the  (a,b)H/V and (c,d)A/D basis for $\tau_p=98$ fs. The interference visibilities and Bell parameters for all our measurements are displayed in Table \ref{tab:bell}. The Bell inequality is clearly violated for all the three different kinds pump setting. This shows that the source always generates entangled photon pairs. Note that there was no need to subtract the H/V background counts.



Finally, we analyze the polarization entangled state produced by the source using quantum state tomography methods. To do that, we placed polarization analyzers consisting of  a polarizer, quarter-wave plate and half wave plate in each arm. We performed 36 coincidence measurements for analyzers set to all combinations from the set of polarizations: H, V, D, A, left (L), right (R) \cite{Thew2002}. Based on those measurements, a standard polarization tomography resulting in the density matrix was performed. An exemplary reconstructed polarization state for a pump pulse duration of $\tau_p=98$ fs is shown in \figref{fig:fringes} (c) and (d). We quantify the quality of the state by computing its fidelity compared to the expected state $(\ket{HH}+\ket{VV})/\sqrt{2}$. We also calculated the entanglement monotone (concurrence) and the purity. The computed values for our measurements are displayed in Tab.~\ref{tab:bell}.


In summary, we have shown that the source characterized previously in Ref.~\cite{Lutz2013} produces both, any desired kind of spectral correlations and high quality polarization entanglement in the same time. This can be used for long distance quantum communication where photons sent through long single mode fibers suffer from chromatic dispersion. The same positive spectral correlation feature has a potential to improve two photon absorption experiments, where two photon coherence effects are important.  

The authors  acknowledge  funding from NSERC (CGS, QuantumWorks, Discovery, USRA), Ontario Ministry of Research and Innovation (ERA program and research infrastructure program), CIFAR, Industry Canada and the CFI, and support by OEC, which provided InGaAs/InP SPAD module produced by Micro-Photon-Devices.  PK acknowledges support by Mobility Plus project no.~602/MOB/2011/0  financed by Polish Ministry of Science and Higher Education and by Fundation for Polish Science under Homing Plus no.~2013-7/9 program supported by European Union under PO IG project. We also thank Zhizhong Yan and Deny Hamel for the fruitful discussions about the NFADs, and Carmelo Scarcela and Alberto Tosi from Politecnico di Milano for insightful discussions about detection techniques.


\begin{thebibliography}{10}
\newcommand{\enquote}[1]{``#1''}

\bibitem{Uren2005}
A.~B. U'Ren, C.~Silberhorn, K.~Banaszek, I.~A. Walmsley, R.~Erdmann, W.~P.
  Grice, and M.~G. Raymer, Las.~ Phys. \textbf{15}, 146 (2005).

\bibitem{Uren2007}
A.~B. U'Ren, Y.~Jeronimo-Moreno, and H.~Garcia-Gracia, Phys. Rev. A
  \textbf{75}, 023810 (2007).

\bibitem{Osorio2008}
C.~I. Osorio, A.~Valencia, and J.~P. Torres, New J. Phys. \textbf{10}, 113012
  (2008).

\bibitem{Mosley2008}
P.~J. Mosley, J.~S. Lundeen, B.~J. Smith, P.~Wasylczyk, A.~B. URen,
  C.~Silberhorn, and I.~A. Walmsley, Phys. Rev. Lett. \textbf{100}, 133601
  (2008).

\bibitem{Kolenderski2009}
P.~Kolenderski, W.~Wasilewski, and K.~Banaszek, Phys. Rev. A \textbf{80},
  013811 (2009).

\bibitem{Eckstein2011}
A.~Eckstein, A.~{Christ}, P.~J. {Mosley}, and C.~{Silberhorn}, Phys. Rev. Lett.
  \textbf{106}, 013603 (2011).

\bibitem{Evans2010}
P.~G. Evans, R.~S. {Bennink}, W.~P. {Grice}, T.~S. {Humble}, and J.~{Schaake},
  Phys. Rev. Lett. \textbf{105}, 253601 (2010).

\bibitem{Gerrits2011}
T.~Gerrits, M.~J. {Stevens}, B.~{Baek}, B.~{Calkins}, A.~{Lita}, S.~{Glancy},
  E.~{Knill}, S.~W. {Nam}, R.~P. {Mirin}, R.~H. {Hadfield}, R.~S. {Bennink},
  W.~P. {Grice}, S.~{Dorenbos}, T.~{Zijlstra}, T.~{Klapwijk}, and V.~{Zwiller},
  Opt. Express \textbf{19}, 24434 (2011).

\bibitem{Hendrych2007}
M.~Hendrych, M.~{Micuda}, and J.~P. {Torres}, Opt. Lett. \textbf{32}, 2339
  (2007).

\bibitem{Valencia2007}
A.~Valencia, A.~Cer\'{e}, X.~Shi, G.~Molina-Terriza, and J.~P. Torres, Phys.
  Rev. Lett. \textbf{99}, 243601 (2007).

\bibitem{Jin2013}
R.-B. Jin, R.~Shimizu, K.~Wakui, H.~Benichi, and M.~Sasaki, Opt. Express
  \textbf{21}, 10659 (2013).

\bibitem{Dayan2005}
B.~Dayan, A.~Pe'Er, A.~A. Friesem, and Y.~Silberberg, Phys. Rev. Lett.
  \textbf{94}, 043602 (2005).

\bibitem{Dayan2007}
B.~Dayan, Phys. Rev. A \textbf{76}, 043813 (2007).

\bibitem{Lutzthes2013}
T.~Lutz, \enquote{Entangled photon sources for ultra-long distance quantum
  entanglement distribution}, Master's thesis, Univeristy of Waterloo, Ulm
  University (2013).

\bibitem{Kuzucu2005}
O.~Kuzucu, M.~{Fiorentino}, M.~A. {Albota}, F.~N. {Wong}, and F.~X.
  {K{\"a}rtner}, Phys. Rev. Lett. \textbf{94}, 083601 (2005).

\bibitem{Kuzucu2008}
O.~Kuzucu, F.~N.~C. Wong, S.~Kurimura, and S.~Tovstonog, Phys. Rev. Lett.
  \textbf{101}, 153602 (2008).

\bibitem{Harder2013}
G.~Harder, V.~Ansari, B.~Brecht, T.~Dirmeier, C.~Marquardt, and C.~Silberhorn,
  Opt. Express \textbf{21}, 13975 (2013).

\bibitem{Lutz2013}
T.~Lutz, P.~Kolenderski, and T.~Jennewein, Opt. Lett. \textbf{38}, 697 (2013).

\bibitem{Kim2002}
Y.-H. Kim and W.~P. Grice, J. Mod. Opt. \textbf{49}, 2309 (2002).

\bibitem{Poh2007}
H.~S. Poh, C.~Y. Lum, I.~Marcikic, A.~Lamas-Linares, and C.~Kurtsiefer, Phys.
  Rev. A \textbf{75}, 043816 (2007).

\bibitem{Tosi2012a}
A.~Tosi, A.~{Della Frera}, A.~{Bahgat Shehata}, and C.~{Scarcella}, Rev. Sci.
  Instrum. \textbf{83}, 013104 (2012).

\bibitem{Thew2002}
R.~T. Thew, K.~{Nemoto}, A.~G. {White}, and W.~J. {Munro}, Phys. Rev. A
  \textbf{66}, 012303 (2002).

\end{thebibliography}
\end{document}